\documentclass[aps,prl,twocolumn,showpacs]{revtex4-1}

\usepackage[dvipdfmx]{color}
\usepackage[dvipdfmx]{graphicx}

\usepackage{dcolumn}
\usepackage{bm}
\usepackage{multirow}

\begin{document}

\title{
Spontaneous Appearance of Low-dimensional Magnetic Electron System on Semiconductor Nanostructures
}

\author{Keisuke Sawada}
\author{Jun-Ichi Iwata}
\author{Atsushi Oshiyama}

\affiliation{
Department of Applied Physics, The University of Tokyo, Tokyo 113-8656, Japan
 }

\date{\today}

\begin{abstract}
We find that spin-polarized ground states emerge in nanofacets which are self-organized on SiC (0001) surfaces.
Our large-scale density-functional calculations reveal that the nanofacet formed by bunching of single bilayer steps generates peculiar carbon dangling bond states localized at but extended along step edges.
The flat-band characteristics of those C states cause either ferromagnetic or anti-ferromagnetic chains on covalent semiconductors.
\end{abstract}


\maketitle

Nanometer-scale structures are now accessible through cutting-edge fabrication technique \cite{itrs} or a gift of self-organization phenomena \cite{shchukin, rousset, giesen, eaglesham, hirayama}. Common expectation that such nanostructures break through limitations of current technology may rely on a fact that nano-scale shapes affect wave-functions of relevant electron states. However, a way of designing nano-shapes which decisively affect the electron states and therefore are intriguing in science and useful in technology is lacking.  

One of the unexpected but promising examples is the spin-polarization or the magnetism in nanostructures of covalent semiconductors, that is occasionally called $d^0$ magnetism \cite{akiyama,uchida,gohda,vankatesan}. Particular arrangements of unpaired electrons on hydrogen-covered Si (111) surface \cite{okada1} and on Au-covered high-Miller index Si surface \cite{erwin1,erwin2} are predicted to be ferrimagnetic and anti-ferromagnetic (AFM), respectively, by the density-functional calculations. The spin-polarization on the Si surface is indeed evidenced by 
photoemission\cite{bierdermann,yeom} and scanning tunneling microscope\cite{aulbach} experiments.

However, such nanostructures are achieved only by sophisticated fabrication techniques on semiconductor surfaces. It would be free from such difficulty if the self-organization could produce desirable nanostructures by itself. We here report occurrence of magnetic chains on self-organized nanostructures of silicon carbide (SiC). 

\begin{figure}
\begin{center}
\hspace*{2mm}
\includegraphics[width=1.25\linewidth] {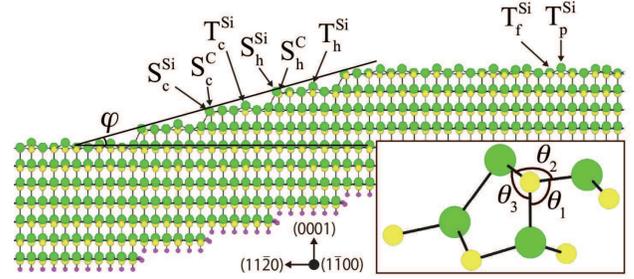}
\caption{\label{structure}
(color online)
Side views of the (11$\bar{2}n$) ($n=12$) nanofacet with the facet angle $\varphi =$ 15.3$^{\circ}$ on the 4H-SiC(0001) 
vicinal surface, which we have found is most stable. Green, yellow and purple balls indicate the Si, C and H atoms, respectively. The T$^{\rm Si}_{\rm p}$ and T$^{\rm Si}_{\rm f}$ indicate the protruded and flat Si sites at the long terrace, respectively. The T$^{\rm Si}_{\rm c}$ and T$^{\rm Si}_{\rm h}$ denote 
the Si sites on the short terraces where the stacking are cubic and hexagonal, respectively. The S$^{\rm Si}_{\rm c}$ or S$^{\rm Si}_{\rm h}$ indicates the Si site at the step edge on the corresponding terrace. 
The S$^{\rm C}_{\rm c}$ and S$^{\rm C}_{\rm h}$ denote the corresponding edge C sites.
The $\theta_1$, $\theta_2$, and $\theta_3$ indicate the bond angles at the step edge C atom.
Expanded view around the step edge and top view of the longest terrace are shown in Supplementary Fig. 1 \cite{SM}.}
\end{center}
\end{figure}

SiC is a wide gap covalent semiconductor, existing as many polytypes such as 2H (wurtzite), 3C (zincblende), 4H, 6H and so forth \cite{polytype}. 
The 4H is the most stable with larger cohesive energy than others by 0.2 meV per SiC unit. 
Due to its high dielectric breakdown voltage and the high melting temperature\cite{SiC}, SiC is a strong candidate for the next-generation devices in environment-friendly power electronics\cite{SiC,rabkowski}. This material shows peculiar nanostructures through self-organization: During the epitaxial growth of SiC (0001) surface, the (11$\bar{2}n$) nanofacets are frequently observed \cite{nakamura,nakagawa,fujii} (Fig.~\ref{structure}); the facet angle $\varphi$ which is the angle between (0001) and (11$\bar{2}n$) is almost the same (magic angle, $\varphi$ = 12$^{\circ}$ - 16$^{\circ}$) and the height of the nanofacet along the (0001) direction is a integer or a half-integer of the unit-cell height\cite{nakagawa,fujii,kimoto,arima,nie}. The mechanism of such nanofacet formation is clarified by our recent density-functional calculations \cite{sawada}: 
Microscopically, the nanofacet is the bunched single bilayer 
atomic steps; the balance between the step-step repulsive energy and the surface-energy gain on the terrace is the reason for the formation of the particular nanofacet structure, strongly indicative of that such nanaofacet is ubiquitous on semiconductor surfaces. 

In this Letter, we show that peculiar electron states which are localized near but extended along step edges of the nanofacets of SiC exhibit 
either ferromagnetic (FM) or AFM spin polarization.
Based on the density-functional calculations, we find that hydrogen passivation of the surface spontaneously induce such spin polarization. 

\begin{figure*}[t]
\begin{center}
\includegraphics[width=0.95\linewidth] {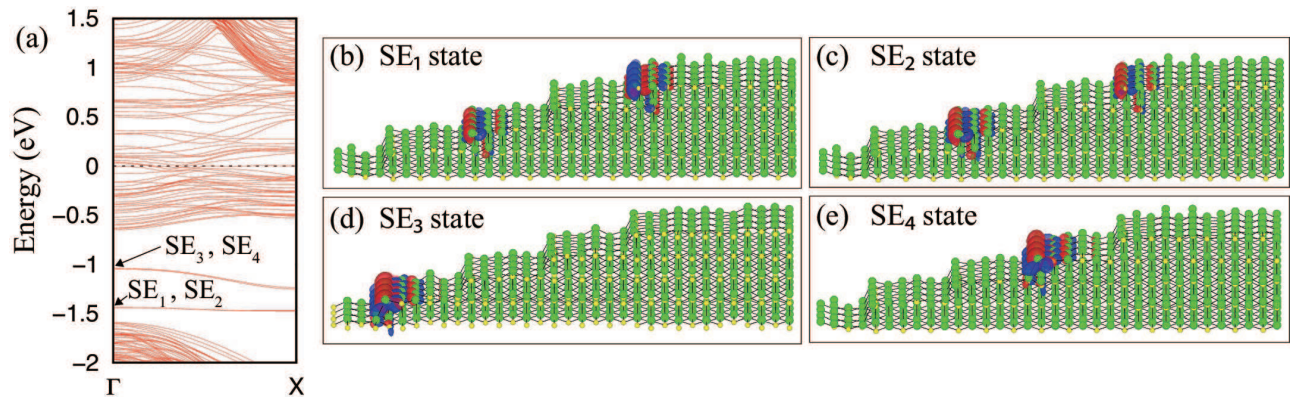}
\caption{\label{band_wf}
(color online)
Energy bands along the $\Gamma$X parallel to the step edge of the stable (11$\bar{2}n$) nanofacet on 4H-SiC (0001) surface obtained by our LDA calculations. The Fermi level is set to be zero. The SE$_1$ and SE$_2$, and also the SE$_3$ and SE$_4$ are almost degenerate. The Kohn-Sham orbitals of the SE$_1$, SE$_2$, SE$_3$ and SE$_4$ states at the $\Gamma$ point are shown in (b), (c), (d), and (e), respectively, as isovalue surfaces. The red and blue colors depict positive and negative values, respectively, of the Kohn-Sham orbitals.
Expanded views of the Kohn-Sham orbital around the step edge are shown in Fig.~\ref{extend_se3}.}
\end{center}
\end{figure*}

Most calculations have been performed by the local density approximation (LDA) \cite{kohn,lda} in the density functional theory (DFT) \cite{hohenberg,kohn}. Key results are crosschecked by the generalized gradient approximation (GGA) \cite{gga} to the exchange-correlation energy functional.
We use our own real-space code named RSDFT \cite{iwata1,iwata2,hasegawa} combined with the norm-conserving pseudo-potentials \cite{TM}.
We have examined effects of core corrections \cite{pcc} on the magnetic states and found that they are unimportant in the system of C and Si.
The real-space scheme is most suitable to current multi-core massively parallel architecture and
allows us to perform extensive calculations for thousands-atom systems. 
Details are given in Supplementary Material \cite{SM}.

We have performed extensive total energy calculations for the (11$\bar{2}n$) nanofacet structures of 4H-SiC \cite{SM} and obtained the most stable structure with the facet angle being 15.3$^{\circ}$ \cite{sawada}, in accord with the experiments \cite{nakagawa,fujii,kimoto,arima,nie}. The structure is a nano-aggregate of bunched single bilayer steps, as shown in Fig.~\ref{structure} and Supplementary Figure 1 (SFig.~1) \cite{SM}. 
The bilayer atomic stacking is ABCBABCB $\cdots$ from the topmost surface. 

\begin{figure}[b]
\begin{center}
\hspace*{10mm}
\includegraphics[width=1.40\linewidth] {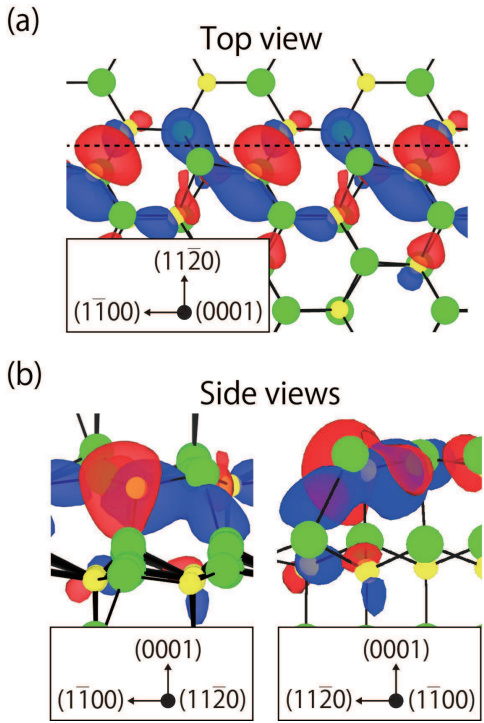}
\caption{\label{extend_se3}
(color online)
Expanded views of the Kohn-Sham orbital of the SE$_3$ state [Fig.~\ref{band_wf}(d)] near the step edge.
The top and side views are displayed in (a) and (b), respectively. 
The dashed line indicates the step edge position.
The color code is the same as in Fig.~\ref{band_wf}. 
}
\end{center}
\end{figure}

Let us start with the calculated electronic structure of this nanofacet (Fig.~\ref{band_wf}). Near the Fermi level ($E_{\rm F}$), we observe many states which have characters of dangling bonds (DBs) of surface Si atoms on the long terrace. The energy bands more than 1.5 eV below $E_{\rm F}$ is mainly ascribed to the valence bands of the bulk SiC, whereas the bands at about 1 eV above $E_{\rm F}$ are from the bulk conduction bands. Apart from those states, we have found four distinct states in the energy gap, labeled as SE$_1$, SE$_2$, SE$_3$ and SE$_4$ in Fig.~\ref{band_wf}(a). From the analyses of the Kohn-Sham (KS) orbitals of these states 
[Figs.~\ref{band_wf} (b)-(e) and Fig.~\ref{extend_se3}], 
we have found that they originate from the step-edge states: The KS orbitals of the SE$_1$ and SE$_2$ states are localized near the step edges where the upper terraces have cubic symmetry (i.e., the stacking sequence from the top 
is either ABC or CBA; cub-terrace edge hereafter), whereas the orbitals of SE$_3$ and SE$_4$ states are near the step edges where the upper terraces have the hexagonal symmetry (i.e. the stacking sequence is either BCB or BAB; hex-terrace edge hereafter). They are localized near but extended along the step edges, exhibiting dispersion-free flat bands.

Figure \ref{extend_se3} clearly shows that those step-edge states have the character of the dangling bonds (DBs) of the three-fold coordinated edge C atoms. They are also mixed with the back-bond character of neighboring edge Si atoms. The states which have the character of DBs of the edge Si atoms are located around $E_{\rm F}$, being substantially resonant with the terrace DB states. Larger electron affinity of C than Si renders the DB-character state of C to be located at lower energy positions being distinct from the Si DB-character states. 

\begin{table}
\begin{center}
\caption{\label{angle} 
Bond angles of the step-edge C atoms in the (11$\bar{2}$n) nanofacet with the upper terraces being of cubic (cub) and hexagonal (hex) symmetry. The $\theta_1$, $\theta_2$, and $\theta_3$ indicate the bond angles shown in Fig. \ref{structure}.}
\begin{tabular}{ccccccc}
\hline 
\hline
& terrace & \ \ \  cub  &  \ \ \ cub &  \ \ \ hex &  \ \ \ hex   \\
\hline
& $\theta_1$ ($^{\rm o}$) & \ \ \ 98.9 & \ \ \ 99.3 & \ \ \ 96.1 & \ \ \ 96.3 & \\
& $\theta_2$ ($^{\rm o}$) & \ \ \ 126.8 & \ \ \ 127.5 & \ \ \ 129.8 & \ \ \ 130.5 & \\
& $\theta_3$ ($^{\rm o}$) & \ \ \ 101.4 & \ \ \ 101.7 & \ \ \ 108.7 & \ \ \ 108.6 & \\
\hline
\hline
\end{tabular}
\end{center}
\end{table}

The splitting between the SE$_1$ and SE$_2$ group and the SE$_3$ and SE$_4$ group comes from structural characteristics of the step edges. Table \ref{angle} shows the bond angles, $\theta_1$, $\theta_2$, and $\theta_3$, at the step-edge C atom. We have found small but sizable difference in the bond angles between cub-terrace edge and hex-terrace edge C atoms. Applying the $\pi$-orbital axis vector analysis (POAV) \cite{hyb}
to the three-fold coordinated C atoms where the three bonds are formed by $sp^n$ hybridization and the remaining DB has a character of $s^mp$ orbital, we can deduce the amount of hybridization $m$ and $n$ from the bond angles, $\theta_1$, $\theta_2$, and $\theta_3$ \cite{hyb}. The obtained values for the amounts of $s$ characters in the DBs are $m = 0.25$ and $m = 0.23$ for the cub-terrace edge C and $m = 0.13$ and $m= 0. 13$ for the hex-terrace edge C. 
The larger mixing with $s$-character renders the SE$_1$ and SE$_2$ states lower than SE$_3$ and SE$_4$ states by a half eV.

\begin{table}[b]
\begin{center}
\caption{\label{adsorption} 
Adsorption energies of hydrogen atom $E_{\rm ad}$ at various Si and C sites depicted in Fig. \ref{structure} of the 4H-SiC nanofacet.
}
\begin{tabular}{ccccccccccc}
\hline
\hline
& site \ & \ T$^{\rm Si}_{\rm f}$ & \ T$^{\rm Si}_{\rm p}$  & 
\ T$^{\rm Si}_{\rm cub}$  & \ T$^{\rm Si}_{\rm hex}$  & 
\ S$^{\rm Si}_{\rm hex}$ & \ S$^{\rm C}_{\rm hex}$ & 
\ S$^{\rm Si}_{\rm cub}$ & \ S$^{\rm C}_{\rm cub}$ & \\ 
\hline
& $E_{\rm ad} \ $
& \ 1.71 & \ 1.68 & \ 1.67 & \ 1.65 & 
\ 1.56 & \ 1.34 & \ 1.33 & \ 1.25 & \\
\hline
\hline
\end{tabular}
\end{center}
\end{table}

Figure \ref{band_wf} unequivocally show the existence of flat bands on semiconductor nanofacets. The flat band may generally causes ferromagnetism \cite{lieb,mielke,tasaki}, and its manifestation in graphene is reported \cite{fujita,okada2}. We now find that hydrogen passivation realizes the possibility of flat-band magnetism 
on the SiC nanofacet. It is of note that the H-passivation on semiconductor surfaces is an extremely common and useful fabrication technique \cite{tsuchida,sieber,seyller}. Table \ref{adsorption} shows the calculated H-adsorption energies $E_{\rm ad}$ \cite{adsorption} for various atomic sites on the nanofacet. 
It is found that $E_{\rm ad}$ depends on the adsorption site. First, the H adsorption with the Si atoms is favorable energetically compared with the C atoms. As stated above, the orbital energy of the Si dangling bond (DB) is higher due to its smaller electron affinity. Before H passivation, some of Si DBs are doubly occupied and others are empty. The H passivation of Si DBs generates doubly occupied Si-H bonding states, thus leading to the larger H adsorption energy. Second, the H adsorption with the edge atoms is less favorable compared with the terrace atoms. This is due to the relaxation around the step edges: As represented in Table \ref{angle}, the edge DBs have more $s$ component in their characters, leading to lower orbital energies and less energy gain upon the H passivation. 
The adsorption energy of the Si atom on the hex-terrace edge (S$^{\rm Si}_{\rm hex}$) is relatively large. This is due to the attractive electrostatic interaction between Si and C in the 4H structure \cite{edgeSi}. 

\begin{figure}
\includegraphics[width=0.9\linewidth] {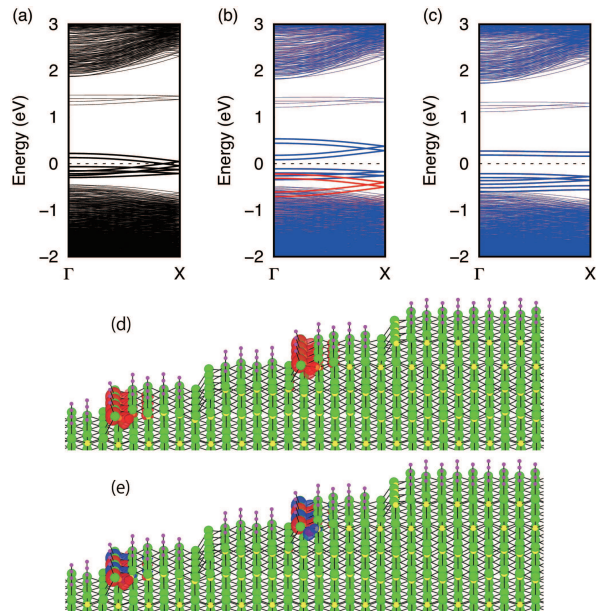}
\caption{\label{spin}
(color online) 
Energy bands [(a)-(c)] and the spin density [(d)-(e)] of the 4H-SiC nanofacet with terrace Si and some of edge Si atoms passivated by H (see text). Energy bands along the step edge ($\Gamma$X) 
for (a) non-magnetic, (b) ferromagnetic (FM) and (c) anti-ferromagnetic (AFM)
states are shown. The Brillouin zone is a half of that in Fig.\ref{band_wf}. 
The energy bands originated from localized states at the C atoms on the cub-terrace and hex-terrace edges are marked by thick lines. 
$E_{\rm F}$ is set to be zero. The red and blue lines in (b) correspond to majority and minority spin states, respectively.
The spin densities in (d) FM and (e) AFM states are shown. 
The red and blue colors in (d) and (e) represent the positive and negative vales for the difference between the up and down spin densities.
Expanded views of spin densities are shown in SFig~2 \cite{SM}.
The color code depicting each atom is same as in Fig.~\ref{structure}.}
\end{figure}

Calculated difference in H-adsorption energies in Table \ref{adsorption} indicates that the terrace Si atoms are easily terminated by the H adsorption, 
whereas the edge atoms are likely to survive with their DBs. The Si atom at the hex-terrace edge is an exception with its large adsorption energy explained above. Hence, when we set an energy criterion of 0.2 eV, it is plausible that the surface atoms are passivated by H except for the edge C atoms and the Si atoms at the cub-terrace edge. This situation is  possible by control of temperature. 

Figure \ref{spin} shows energy bands and spin densities of such H-passivated nanofacet. 
We have found that $E_{\rm F}$ shifts downward and crosses the flat bands \cite{passivation}. Then we have obtained three solutions of KS equation: 
The non-magnetic (NM), FM and AFM states.
The total energy of the AFM state is the lowest, being followed by the FM state with the energy increase of 19 meV (7.9 meV) per spin and then the NM state with further increase of 83 meV (225 meV) in our LDA (GGA) calculations.
In the NM state, $E_{\rm F}$ crosses the energy bands with the DB character of the hex-terrace edge C atoms (states SE$_3$ and SE$_4$). The character of Si DBs at the edge is absent in these bands due to the electron affinity difference 
so that the dispersion along the edge becomes smaller than expected from the atomic arrangement. Hence the exchange plus correlation splitting in the KS orbitals of the C DBs is 0.78 eV and 0.72 eV in the FM and AFM states, respectively, leading to the spin-polarized state in the SiC nanafacet. The AFM state in preference to the FM state which we have found is provisionally ascribed to 
the prominent flat-band nature in the AFM state induced by the localization of the edge carbon DB (CDB) state. 
However, it is plausible that the AFM and FM states coexist in suitable temperature range. It is of note that the spin density [Figs.~\ref{spin} (d) and (e)] is solely distributed along the hex-terrace edges, reflecting atom-scale reconstruction near edges explained above. 

\begin{figure}
\includegraphics[width=0.9\linewidth] {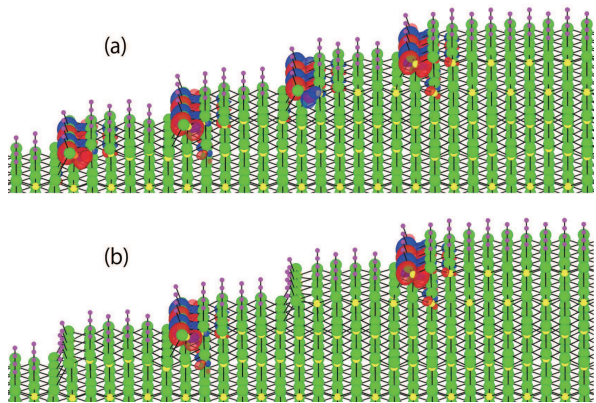}
\caption{\label{spin2}
(color online)
Spin densities of the 4H-SiC nanofacet with terrace Si atoms and edge Si atoms passivated (a), and further hex-terrace edge C atoms additionally passivated (b). 
The color code is the same as in Fig.~\ref{spin}.
Expanded views are shown in SFig.~3 \cite{SM}.
}
\end{figure}

The situation we have explained is characterized as the spin-polarization in the three-quarter filling flat bands. The filling by electrons is controlled by the H-passivation. When all the edge atoms are free from the H-passivation, which may be possible by increasing temperature, we have a situation that all CDB states are occupied 
and Si DB states are unoccupied, leading to the semiconducting character (not shown here).
When all the Si edge atoms as well as the terrace Si atoms are passivated, 
the CDB states are half-occupied. We have found the AFM and FM states in this case in the range of the total energy of 9.9 meV. 
The NM state is higher in energy by 100 meV. The spin density in the AFM state is shown in Fig.~\ref{spin2} (a). Interestingly, the spin density is now distributed to all the edges including the cub-terrace edges. This is caused by the filling control of the CDB states by the H passivation. When the C atoms at the cub-terrace edge, which produce the least adsorption energy, is free from the H passivation, we encounter a situation in which a CDB state at the cube-terrace edge is half filled. We have found the FM and AFM states in the range of 0.1 meV per spin, being lower than the NM state by 141 meV. In this case, the spin density is distributed solely along the cub-terrace edges[Fig.~\ref{spin2}(b)].


In summary, we have performed large-scale density-functional calculations on the stable 4H-SiC nanofacet and found that the spin polarization along the step edges in the nanofacet emerges. This comes from the localization of the electron states caused by the electron affinity difference between C and Si and by the peculiar atomic structures of SiC surface. 
We have found that the hydrogen passivation is a powerful tool to control the spin polarization in the system with ubiquitous elements of H, C and Si.

\begin{acknowledgments}
This work was supported by the Grants-in-Aid for scientific research under Contract No. 22104005 and also by ``Computational Materials Science Initiative", both conducted by MEXT, Japan.  Computations were performed mainly at K Computer in Advanced Institute for Computational Science, RIKEN and at Supercomputer Center in ISSP, University of Tokyo.
\end{acknowledgments}


\begin{thebibliography}{99}
\bibitem{polytype} 
In tetrahedrally bonded semiconductors, there are possible polytypes in which the stacking of atomic bilayers along the bond direction is different. Each polytype is labeled by the periodicity of the stacking sequence and the symmetry (cubic or hexagonal) such as 2H (wurtzite), 3C (zincblende), 4H, and 6H. The sequence of the most stable 4H polytype is $\cdots$ ABCB $\cdots$ where A, B and C denote 3 possible ways of the stacking.
\bibitem{adsorption}
The H-adsorption energy is defined as $ E_{\rm ad} = (E_{\rm NF} + \mu_{\rm H}) - E_{\rm HNF} $ with $E_{\rm NF}$ and $E_{\rm HNF}$ being the total energies of the nanofacet without and with, respectively, H atoms. The $\mu_{\rm H}$ is the chemical potential of H atoms and is regarded here as a half of an energy of an H$_2$ molecule.
\bibitem{edgeSi}
Due to the difference in electronegativity, the electron transfer from Si to C takes place in SiC. At the site S$^{\rm Si}_{\rm hex}$, the third-nearest-neighbor C atom in the subsurface is located with the short distance of 2.86\AA. This leads to the electrostatic energy gain upon H adsorption which causes positively charged Si edge atom.
\bibitem{passivation}
Hydrogen passivation of Si annihilates Si DB states, which are either unoccupied or doubly occupied, and induces H-Si bonding and anti-bonding states, thus shifting $E_{\rm F}$ downwards.
\end{thebibliography}
\end{document}


\title{\bf 
Supplementary Material for: \\
Spontaneous Appearance of Low-dimensional Magnetic Electron System on Semiconductor Nanostructures}
\author{Keisuke Sawada, Jun-Ichi Iwata, and Atsushi Oshiyama}
\affil{\normalsize \it Department of Applied Physics, The University of Tokyo, Tokyo 113-8656, Japan}
\date{}
\maketitle

\section*{Details of Calculations}

To perform accurate large-scale total-energy electronic-structure calculations efficiently on the current and future architectures,
we have adopted a real-space (RS) scheme\cite{chelikowsky,iwata1} in the density functional theory (DFT) combined with the first-principles pseudo-potential scheme, and have developed a highly efficient computation code named RSDFT\cite{iwata2}.
In the real-space scheme, discrete grid points are introduced in real space,
and the Hamiltonian in the Kohn-Sham equation in the DFT is expressed as a matrix represented in terms of the grid-point space.
Differential operators for kinetic energy are replaced by finite-difference operators with sufficiently high orders.
Hence, the Hamiltonian matrix becomes sparse and Fast-Fourier Transformation (FFT) is rarely required for the Hamiltonian operation.
FFT generally causes heavy communication burden among all the compute nodes in the conventional plane-wave basis-set scheme.
As a result, the communication in the RSDFT code arises from the finite-difference operations, the non-local pseudo-potential operations, and the non-local exchange-correlation functional operations. 
Hence, the communication becomes local only among neighboring compute nodes so that the total communication cost is tremendously reduced.
In fact, the RSDFT shows good scalability even with 80,000 compute nodes at K computer at Kobe, Japan\cite{hasegawa1,hasegawa2}.

To check the validity of the results, we construct a periodic single-bilayer-step (SBS) structure without a long terrace. 
In the present calculations, the grid spacing is taken to be 0.22 {\AA} corresponding to a cutoff energy of 57 Ry in the plane-wave-basis-set calculations.
The reliability for this grid spacing is ensured by the calculations with the grid spacing 0.17 {\AA}, corresponding to a cutoff energy of 100 Ry in the plane-wave-basis-set calculations, in the SBS structure.

For the nanofacet formed on 4H-SiC(0001) vicinal surfaces, we treat the Si face of the (0001) surface on which the nanofacets are observed\cite{nakagawa,fujii}. To simulate the surface, we adopt the repeating slab model. 
We use an inclined slab representing the (11$\bar{2}N$) ($N = 32$) vicinal surface (vicinal angle: 5.9$^{\circ}$) and containing single-bilayer steps. 
The bottom surface of the slab is passivated by H atoms. Each slab consists of 7 SiC bilayers and separated from each other along the (11$\bar{2}N$) ($N = 32$) direction by the vacuum region of 10 {\AA}.
The unit length along the step-edge direction $a_0$ for the 1 $\times$ 1 lateral periodicity is taken to be 5.27 {\AA} for LDA and 5.35 {\AA} for GGA, which is the lattice constant obtained by the total-energy calculation for the crystalline 4H-SiC. 
Brillouin-zone (BZ) integration for this slab system is performed by the four $k$ point sampling along the step-edge direction:
the total energy of the SBS structure varies within only 3 meV/cell when ten $k$ points are used.
The geometry optimization is carried out until the remaining forces become less than 50 meV/{\AA}.
The positions of the bottom SiC bilayer passivated by H atoms are fixed throughout the geometry optimization. 
For spin-polarized calculations, in order to explore a possibility of anti-ferromagnetic couplings along step-edge directions,
we take a double periodic cell for the step-edge direction and use two $k$ points along the step-edge direction for BZ integration.
We confirmed that two $k$ points are enough to describe the magnetic stability by comparing the result with that obtained using eight $k$ points.
The geometry optimization in spin-polarized systems is performed by the same condition as the non-spin-polarized calculation.

\section*{Supplementary Figures}

\begin{figure} [htbp]
\begin{center}
\includegraphics[width=0.95\linewidth]{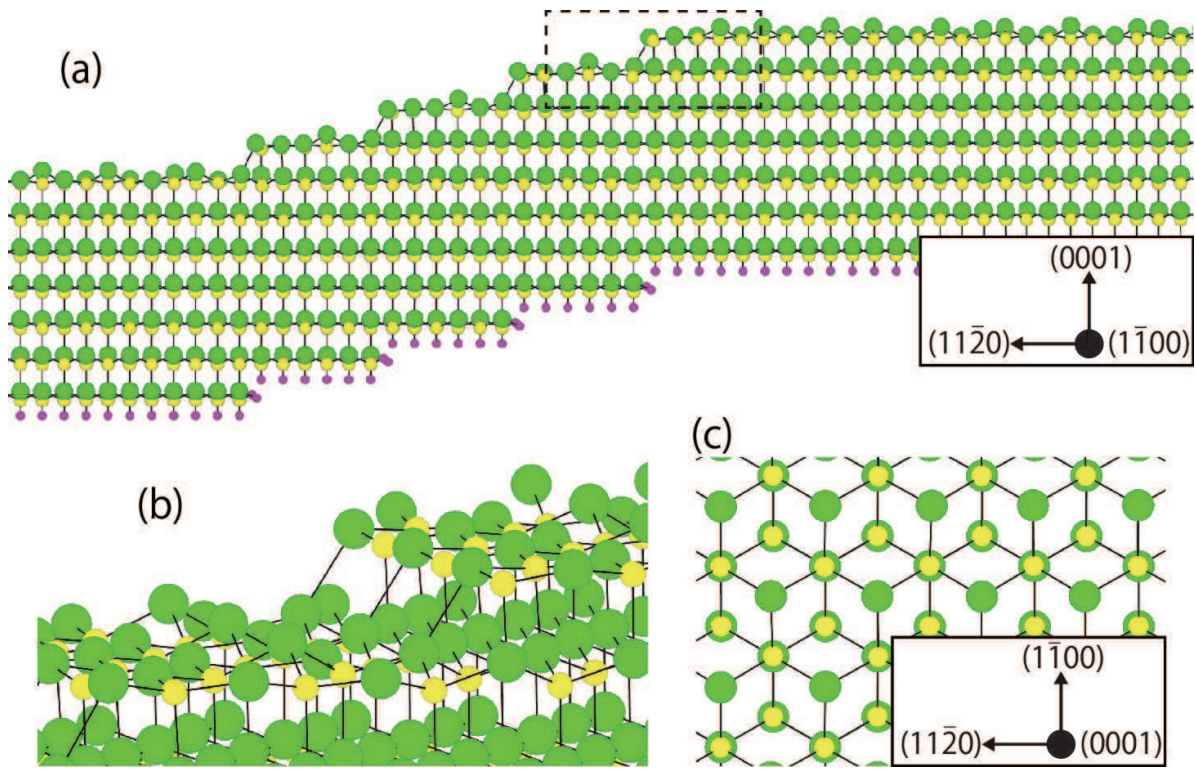}
\caption{(color online)
(a) Side views of the (11$\bar{2}n$) ($n=12$) nanofacet with the facet angle $\varphi =$ 15.3$^{\circ}$ on the 4H-SiC(0001)
vicinal surface, which we have found is most stable. Green, yellow and purple balls indicate the Si, C and H atoms, respectively.
Expanded view around the step edge [dashed square in (a)] and the top view of the longest terrace are displayed in (b) and (c), respectively.}
\end{center}
\end{figure}

\clearpage

\begin{figure} [htbp]
\begin{center}
\includegraphics[width=0.95\linewidth]{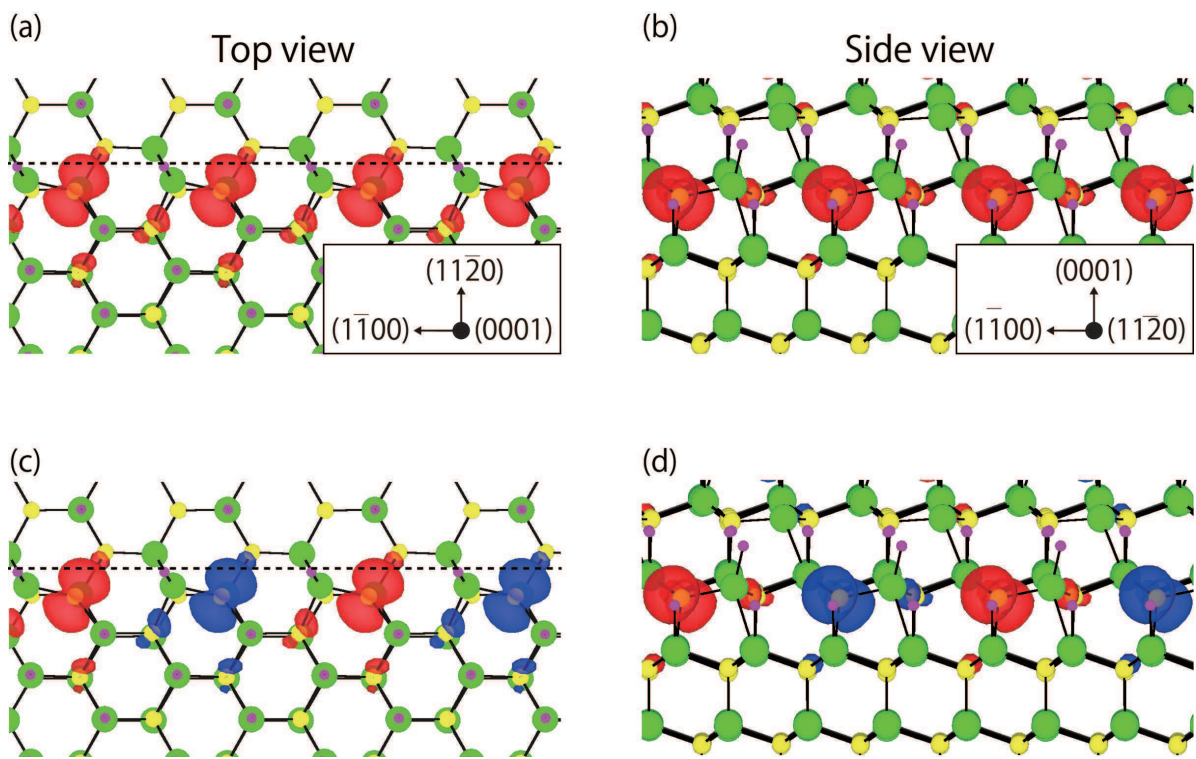}
\caption{(color online)
Expanded views around the step edge for spin densities of the 4H-SiC nanofacet with the H-passivation except for the edge C atoms and the Si atoms at the cub-terrace edge.
The ferromagnetic state [(a) and (b)] and anti-ferromagnetic state [(c) and (d)] are shown.
The dashed lines indicate the step edge positions.
The red and blue colors represent the positive and negative vales for the difference between the up and down spin densities.}
\end{center}
\end{figure}

\clearpage

\begin{figure} [htbp]
\begin{center}
\includegraphics[width=0.95\linewidth]{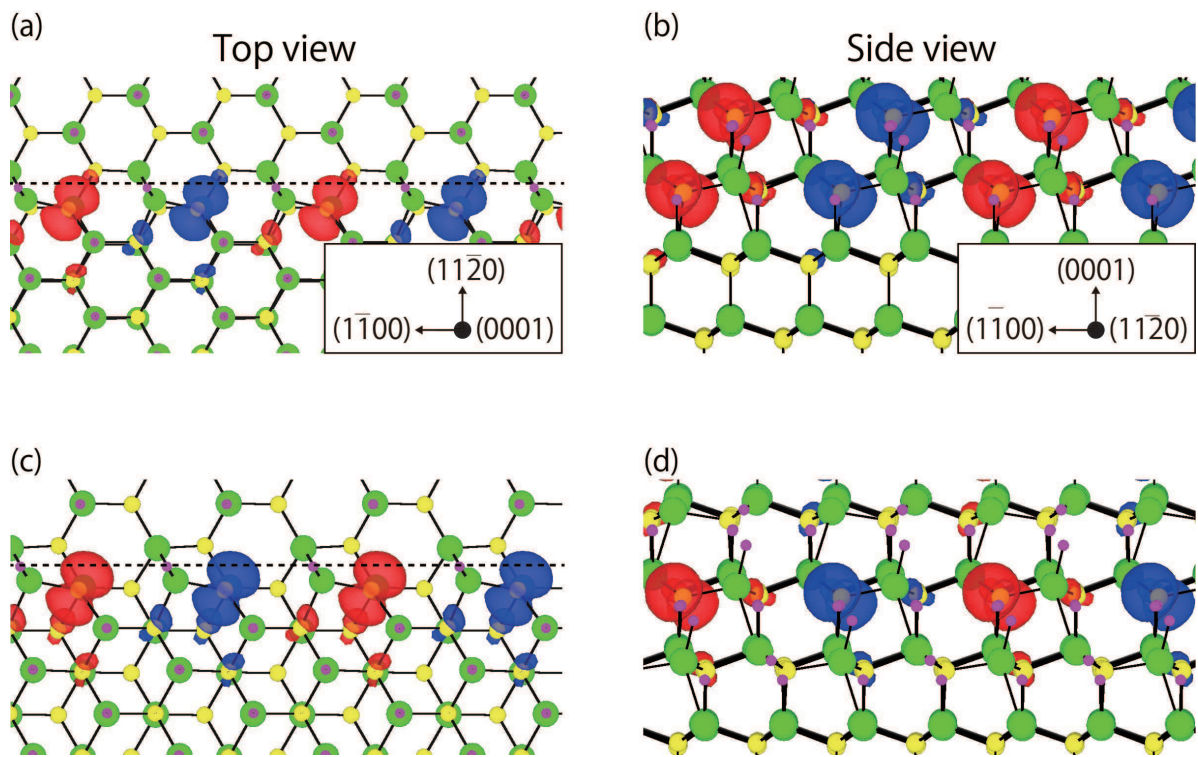}
\caption{(color online)
Expanded views near the step edge for spin densities of the 4H-SiC nanofacet with terrace Si atoms and edge Si atoms passivated [(a) and (b)] and further hex-terrace edge C atoms additionally passivated [(c) and (d)].
The dashed lines indicate the step edge positions.
The red and blue colors represent the positive and negative vales for the difference between the up and down spin densities.}
\end{center}
\end{figure}